\documentclass[preprint,showpacs,nofootinbib,prc,aps,unsortedaddress]{revtex4-1}

\usepackage{graphicx,amstext,amssymb}

\begin{document}

\title{ Thermal nature of charmonium transverse momentum spectra from Au-Au collisions at the highest
energies available at the BNL Relativistic Heavy Ion Collider
(RHIC)}

\author{S.V. Akkelin}
\affiliation{Bogolyubov Institute for Theoretical Physics,
Metrolohichna str. 14b, 03680 Kiev,  Ukraine}

\author{P. Braun-Munzinger}
\affiliation{GSI Helmholtzzentrum f\"ur Schwerionenforschung GmbH,
D-64291 Darmstadt, Germany} \affiliation{ExtreMe Matter Institute
EMMI, GSI, D-64291 Darmstadt, Germany}\affiliation{Technical
University Darmstadt, D-64289 Darmstadt, Germany}
\affiliation{Frankfurt Institute for Advanced Studies, J.W. Goethe
University, D-60438 Frankfurt, Germany}

\author{Yu.M. Sinyukov}
\affiliation {Bogolyubov Institute for Theoretical Physics,
Metrolohichna str. 14b, 03680 Kiev,   Ukraine}

\begin{abstract}
We analyze the transverse momentum distribution of $J/\psi$  mesons
produced in Au + Au collisions at the top RHIC energy within a
blast-wave model that accounts for a possible inhomogeneity of the
charmonium distribution and/or flow fluctuations. The results imply
that the transverse momentum spectra of$J/\psi$, $\phi$ and $\Omega$
hadrons measured at the RHIC can be described well if kinetic
freeze-out takes place just after chemical freeze-out for these
particles.
\end{abstract}


\maketitle

\section{Introduction}

Charmonium production is considered as  important messenger of the
deconfinement transition in heavy-ion collisions. In the original
investigation the suppression of charmonium was proposed as a signal
of the quark gluon-plasma (QGP) that caused the melting of
charmonium \cite{Satz1} (for very recent reviews see Refs.
\cite{Satz2,Rapp0}). Alternatively, the charmonium yield is
described as been caused by  statistical hadronization at the phase
boundary \cite{PBM1}.\footnote{Possible thermal features in
charmonium and, in particular, $J/\psi$ production in
\textit{Pb}+\textit{Pb} collisions at SPS energy were first pointed
out in Ref. \cite{GG}. However, thermal production of charm quarks
is negligible at RHIC energy and below because of the large charm
quark mass \cite{PBM1}.} In the corresponding statistical
hadronization model (SHM), the $J/\psi$ multiplicities at SPS and
RHIC energies are well described for different centrality and
rapidity bins; the basic points of the SHM and its current
development are presented in reviews \cite{PBM2,PBM3} and references
therein. The SHM assumes that the charm quarks are produced in
primary hard collisions and that their total number stays constant
until hadronization. A crucial hypothesis   of the model is  thermal
equilibration of charm quarks in the QGP, at least near the critical
temperature $T_{\text{c}}$. The charmonia are then all produced near
the phase boundary.\footnote{Perhaps slightly above $T_{\text{c}}$:
the upper limit for the dissociation temperature was recently
estimated as $1.2T_{\text{c}}$ \cite{Petreczky}.}

Because the momentum spectrum of $J/\psi$ is nearly frozen out at
$T_c$ (the typical cross section of $J/\psi$ with comoving hadrons
is at most a few millibarns  \cite{Brat}), the measured momentum
spectra of $J/\psi$ mesons contain valuable information on details
of the hadronization process in the strongly interacting medium. For
instance, if formation of $J/\psi$ proceeds as assumed in the SHM,
then the momentum spectrum of $J/\psi$ is expected to be locally
equilibrated in the vicinity of $T_{\text{c}}$. On the contrary, if
the superdense matter formed in \textit{A}+\textit{A} reactions
contained a significant portion of $J/\psi$ mesons produced at the
early stage of the process in hard nucleon-nucleon collisions, then
the final $J/\psi$ spectra could  exhibit two components, as, for
example, in the model \cite{Rapp-2}: direct $J/\psi$ and secondary
thermal $J/\psi$ produced at the stage of recombination of charm
quarks. The shape of $J/\psi$ momentum spectra  will then be, in
fact, nonthermal. Also, in the latter approach, the multiplicities
of hadrons with open and hidden charm do not coincide, generally,
with those predicted in the SHM.

 Recently the PHENIX Collaboration published
results for  $J/\psi$ production versus centrality, transverse
momentum and rapidity in \textit{A}+\textit{A} collisions at
$\sqrt{s_{NN}}=200$ GeV \cite{Phenix}.  The analysis of these data
in Ref. \cite{PBM2} by means of the hydroinspired blast-wave model
\cite{Heinz} demonstrated that a simple blast-wave model
parametrization of the freeze-out does not fully account for the
results  measured by the PHENIX Collaboration on $J/\psi$ transverse
momentum spectra despite the success of the SHM in the description
of multiplicities.  On the contrary, the fit based on the
two-component model \cite{Rapp-2} with the blast-wave
parametrization of transverse momentum spectra for secondary
$J/\psi$ describes well the momentum  spectra of $J/\psi$
\cite{Rapp3}. However, the portion of primordial $J/\psi$ that were
not melted in the QGP seems to be too big in this model: about
$50\%$. Note that a good description of $J/\psi$ spectra can  also
be reached without the assumption of thermal equilibration of the
secondary charmonia. If recombination of charm quarks is based on
the coalescence mechanism (for review see, e.g., Ref. \cite{Greco}
and references therein), then the secondary $J/\psi$ are not thermal
even if charm quarks are considered to be thermalized ones
\cite{Rapp00} at the hadronization stage. The latter scenario is
realized, for example, in Ref. \cite{Rapp1} where the recombination
coalescence component of $J/\psi$ spectra was calculated based on
the kinetic coalescence model of  Ref. \cite{Rapp2}.  However,
although the recombination contribution matches  the low $p_{T}$
region well \cite{Rapp1}, significant primordial $J/\psi$ production
is again required  to describe the higher $p_{T}$
region.\footnote{See also Ref. \cite{nxu}, where the $J/\psi$
transverse momentum distribution was calculated in a transport model
with both initial production and continuous regeneration of
charmonia.}

In  the present article we advocate the local equilibrium picture of
$J/\psi$ production at the freeze-out stage near $T_{\text{c}}$ in
accordance with the statistical hadronization  model
\cite{PBM1,PBM2,PBM3}. While multiplicities in this model were
described in Refs. \cite{PBM2,PBM3}, our aim here is to demonstrate
that transverse momentum spectrum slopes of $J/\psi$ mesons at
midrapidity can be described by a hydroinspired parametrization
without a significant primordial $J/\psi$ contribution: the latter
is questionable taking into consideration recent results
\cite{Petreczky} as for $J/\psi$ melting in the QGP.  The article is
organized as follows. In Sec. II, we present the results of
blast-wave model fits to describe transverse momentum spectra of
$\phi$ and $\Omega$ hadrons at RHIC and discuss the problems with
the fit of $J/\psi$ spectra.  In Sec. III, we briefly analyze the
main assumptions of the blast-wave model, to clear up the reasons
for the shortcomings in the blast-wave model description of $J/\psi$
transverse momentum spectra,  and show that a  consistent
description of the $\phi$, $\Omega$, and $J/\psi$ transverse
momentum spectra can be obtained within a hydroinspired
parametrization of transverse momentum spectra  that accounts for a
possible inhomogeneity of charmonium distribution and/or flow
fluctuations. We conclude in Sec. IV.

\section{Blast-wave model fits to transverse momentum spectra  of
$\phi$ and $\Omega$ at RHIC and transverse momentum spectra  of
$J/\psi$ }

Let us start with the  assumption that the momentum spectra of
$J/\psi$, as well as $\phi$ and $\Omega$ hadrons, are frozen near
the phase boundary in RHIC-energy collisions at  $T=160$ MeV. This
value  is close to the temperature of chemical freeze-out in these
collisions \cite{PBM4,Star-0}. We also note that $T_{\text{c}}$ as
obtained for zero-net-baryon-density matter from the most recent
lattice QCD calculations is close to this value \cite{karsch-10}. We
choose $\phi$ and $\Omega$ particles for comparison with $J/\psi$
mesons because  these particles are thought to have very small
hadronic cross sections and their spectra are not expected to be
significantly  distorted by feed-down from resonance decays in
relativistic heavy-ion collisions. The assumption of coincidence of
the kinetic freeze-out of these particles  with their chemical
freeze-out was supported in Ref. \cite{Barannikova}  where it was
demonstrated that the  blast-wave model fit to the transverse mass
spectra of these particles at RHIC yields $T \approx 160$ MeV along
with an average flow velocity $\langle v \rangle \approx 0.45c$. The
latter is lower than that for hadrons strongly  interacting in the
hadron resonance gas, and, thereby their chemical freeze-out
precedes kinetic freeze-out.\footnote{A similar conclusion was also
obtained in Ref. \cite{Gor} for SPS $158A$ GeV energy collisions
where blast-wave fits to $J/\psi$ and $\psi'$ mesons and $\Omega$
hyperon spectra in \textit{Pb}+\textit{Pb} collisions  were
performed to show that a good description of these particle spectra
by the blast-wave formula with the temperature associated with the
chemical freeze-out and a relatively low transverse collective
velocity can be obtained.} Coincidence of thermal and chemical
freeze-out also underlies the hydroinspired model presented in Ref.
\cite{broniowski}.

It is noteworthy  that the inverse slope parameters for $\Omega$ and
$\phi$ transverse momentum spectra do not vary significantly with
collision centrality \cite{Star1,Phenix1,Star2}.  This weak
centrality dependence can be explained by the independence of the
chemical freeze-out temperature from the  collision centrality that
was observed in the chemical equilibrium model fit to particle
number yields \cite{Star3,Star-0}. This experimental observation
also supports the coincidence of kinetic and chemical freeze-out for
these particle species. Note that the insensitivity of the chemical
freeze-out temperature to centrality is natural \cite{Heinz1} if
chemical freeze-out happens at the phase boundary \cite{PBM5} (see
also the recent review article \cite{PBM6}).

To calculate transverse momentum spectra of $J/\psi$, $\phi$ and
$\Omega$ particles in central collisions we utilize the blast-wave
formula \cite{Heinz},
\begin{eqnarray}
 \frac{dN}{p_{T}dp_{T}dy}\propto m_{T}
\int_{0}^{R} dr r  I_{0}\left( \frac{p_{T}\sinh y_{T}(r)}{T}\right)
K_{1}\left( \frac{m_{T}\cosh y_{T}(r)}{T}\right), \label{01}
\end{eqnarray}
which follows from boost-invariant parametrizations of freeze-out at
$\tau = \sqrt{t^{2} - z^{2}}=const$ under the assumption of
homogeneous particle number density until $r=R$, where $R$ is the
transverse size of the system. Here $I_0$ and $K_1$ represent
modified Bessel functions, and $m_{T}=\sqrt{p_{T}^{2}+m^{2}}$. We
utilize a constant temperature $T$ across the freeze-out
hypersurface and a linear transverse rapidity profile, $y_{T}=
y_{T}^{\text{max}} r/R$, where $y_{T}^{\text{max}}=y_{T}(R)$ is the
maximum transverse rapidity. The collective radial expansion
velocity $v$ is then $v=\tanh y_T$, and $v_{\text{max}}=\tanh
y_{T}^{\text{max}}$ is the velocity at the surface.

Because we are interested  in the description of the transverse
spectrum slope, it is convenient to substitute $r/R$ with $x$ in Eq.
(\ref{01})  to obtain
\begin{eqnarray}
 \frac{dN}{p_{T}dp_{T}dy}\propto m_{T}
\int_{0}^{1} xdx  I_{0}\left( \frac{p_{T}\sinh (\alpha
x)}{T}\right) K_{1}\left( \frac{m_{T}\cosh (\alpha x)}{T}\right),
\label{1}
\end{eqnarray}
with $\alpha=y_{T}^{\text{max}}$.

\begin{figure}[h]
\includegraphics[scale=0.7]{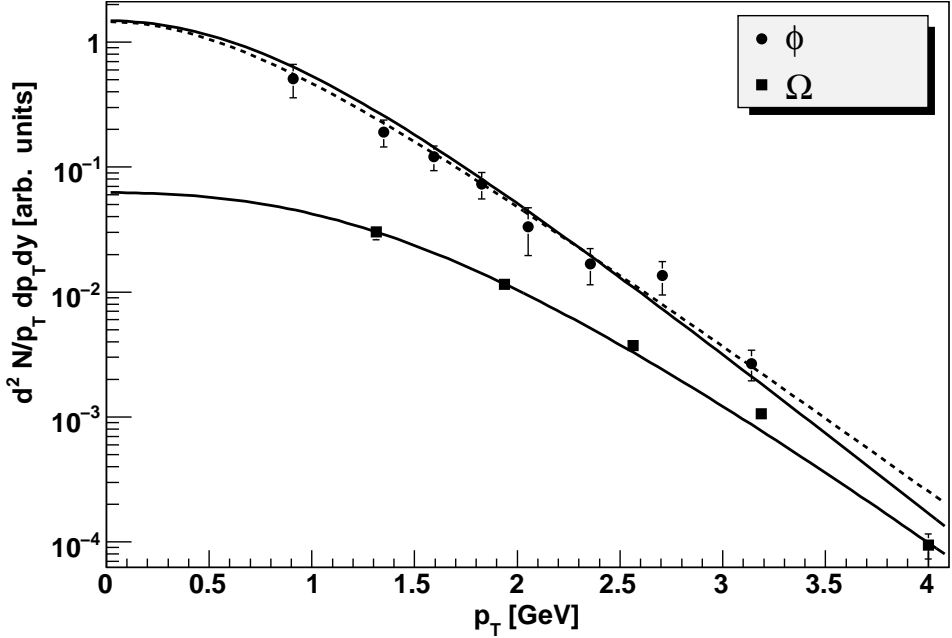}
\caption{\label{figg1}Transverse momentum data (in arbitrary units)
at midrapidity of $\phi$ mesons  and $\Omega $
($\overline{\Omega}^{+}+\Omega^{-}$) hyperons  in central $Au+Au$
collisions at $\sqrt{s_{NN}}=200$ GeV RHIC energy along with the
results of the blast-wave model, Eq.  (\ref{1}),  fit which are
indicated by the solid lines, the dashed line represents the
exponential,  Eq. (\ref{02}), fit to momentum spectra of $\phi$
mesons by the STAR Collaboration \cite{Star2}, which we normalized
to the PHENIX data points. Experimental data for $\phi$ and $\Omega$
are from Refs. \cite{Phenix1} and \cite{Star1} respectively.
Statistical error bars are shown for $\phi$ mesons, and total
(statistical and systematic) error bars are presented for $\Omega $
hyperons.}
\end{figure}

We check first that assuming equal chemical and kinetic freeze-out
temperatures is a good approximation for $\phi$
 and $\Omega$ ($\Omega \equiv \overline{\Omega}^{+} +\Omega^{-}$) particle
transverse momentum spectra at midrapidity. The corresponding
experimental data for \textit{Au}+\textit{Au } collisions at
$\sqrt{s_{NN}}=200$ GeV for $\Omega$ baryons in the $0\%-10\%$
centrality bin are taken from Ref. \cite{Star1} (STAR
Collaboration), and those for $\phi$ mesons at $0\%-10\%$ centrality
from Ref. \cite{Phenix1} (PHENIX Collaboration). As shown in Fig.
\ref{figg1}, the shapes of the $\Omega$ and $\phi$ spectra are
reproduced well by the blast-wave formula with fairly realistic
parameters at chemical freeze-out: $T=160$ MeV, $\alpha = 0.75$,
implying $v_{\text{max}} \approx 0.635$.\footnote{Note that
relatively high collective transverse velocities  for $T=160$ MeV
can appear in the course of the evolution  \cite{flow} as a result
of the development  of initial transverse velocities at the
prethermal partonic stage (with subsequent rapid thermalization) and
viscous effects. The latter leads to a decrease in the longitudinal
velocity and increase in transverse velocity.}
We also show in Fig. \ref{figg1} the exponential fit to momentum
spectra of $\phi$ mesons  in the $0\%-10\%$ centrality bin by the
STAR Collaboration \cite{Star2}, which we normalized to PHENIX data
points. The STAR Collaboration demonstrated that $\phi$-meson
momentum spectra are fitted well by an exponential function,
\begin{eqnarray}
\frac{d^{2}N}{p_{T}dp_{T}dy} \sim \exp{(-m_{T}/T_{\text{exp}})}
\label{02}
\end{eqnarray}
where  the slope parameter $T_{\text{exp}} \approx 359$ MeV for the
$0\%-10\%$ centrality bin. Note that the $dN/dy$ results for $\phi$
mesons extracted  by the PHENIX and STAR Collaborations in this
centrality bin
 are not in agreement, while, as  explicitly shown  in Fig. \ref{figg1}, there is no
discrepancy between the measured spectrum slopes.

\begin{figure}[h]
\includegraphics[scale=0.7]{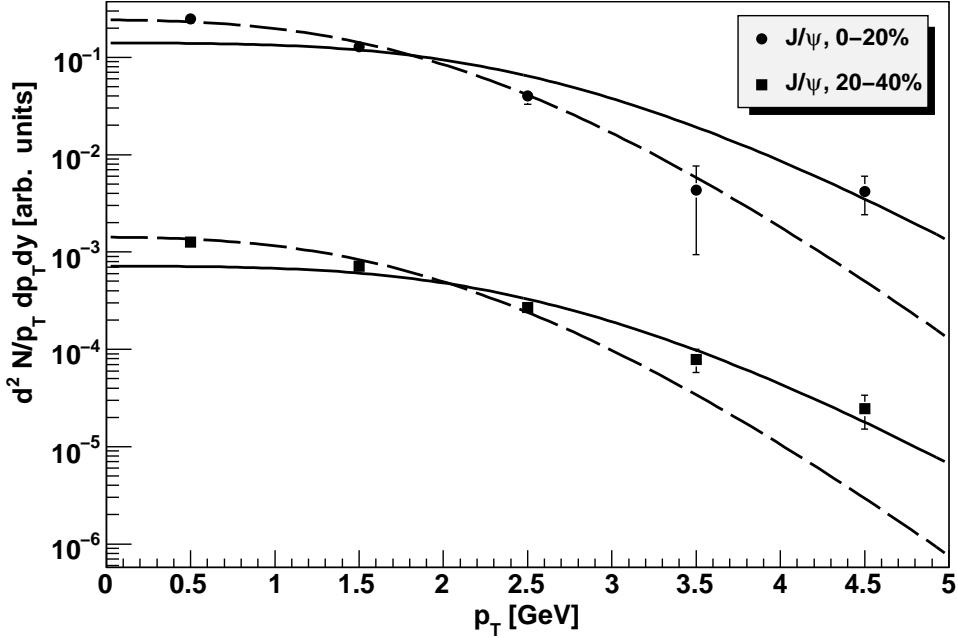}
\caption{\label{fig2} Transverse momentum data for $J/\psi$ mesons
(in arbitrary units) at midrapidity   in  $Au+Au$ collisions at
$\sqrt{s_{NN}}=200$ GeV RHIC energy along with the results of the
blast-wave model, Eq.  (\ref{1}), fit  with the same parameters as
in Fig. \ref{figg1} (solid lines). Dashed lines represent the
blast-wave model, Eq. (\ref{1}), fit results with the "soft"
transverse flow parameter: see text for details. Experimental data
for $J/\psi$ are from Ref. \cite{Phenix} for two centrality bins.
Statistical and systematic uncorrelated errors are represented as
error bars.}
\end{figure}

  The results for  momentum
spectra of $J/\psi$ with the same set of parameters are shown in
Fig. \ref{fig2}, together with the data measured by the PHENIX
Collaboration \cite{Phenix} for two centrality bins.\footnote{We
show the data points of noncentral collisions ($20\%-40\%$
centrality bin) just  for comparison.} One notices significant
deviations for the data points at low momentum, leading to a
excessively  high effective temperature.

To see what  blast-wave model parameters are required to describe
the low-$p_T$ region of $J/\psi$ spectra, we performed a blast-wave
model fit to the low-momentum part of $J/\psi$ spectra at the same
temperature, $T=160$ MeV, but considered the maximum transverse
rapidity $\alpha$ in  Eq. (\ref{1}) to be a  free parameter. The
results are demonstrated  in Fig. \ref{fig2}, where one can see that
the blast-wave model with $\alpha = 0.55$ (implying
$v_{\text{max}}\approx 0.5$) yields a good description of the
low-momentum part of the $J/\psi$ spectrum. These findings, at first
sight, might imply a contradiction with the  blast-wave model fits
to $\phi$ and $\Omega$ transverse momentum spectra where the same
temperature but another surface velocity was obtained. Consequently,
the question arises: Can $J/\psi$, $\phi$ and $\Omega$ transverse
momentum spectra be coherently described within a hydroinspired
parametrization of freeze-out with reasonable parameters?

\section{ Hydroinspired parametrization of $J/\psi$ transverse momentum
spectrum freeze-out at the RHIC}

To understand why a blast-wave model, Eq. (\ref{01}), description
with parameters fitted to $\phi$ and $\Omega$ transverse momentum
spectra does not reproduce  the $J/\psi$ spectra properly, let us
look carefully over the main assumptions used in the comparison to
spectra by employing Eq. (\ref{01}). They are the following: the
freeze-out isotherm is $\tau=const$ and the transverse rapidity
profile at the isotherm is linear. Further assumptions  are on-mass
shell distribution functions, absence of  flow fluctuations, a
homogeneous particle number density, $n=\sqrt{n^{\mu}n_{\mu}}$, at
freeze-out, and absence of  resonance feed-down. Here we briefly
discuss these assumptions to clarify whether giving up debatable
ones can result in consistent description of $\phi$, $\Omega$, and
$J/\psi$ transverse momentum spectra if kinetic freeze-out takes
place just after chemical freeze-out for these particles.

First note that a hypersurface of constant proper Bjorken time,
$\tau = \sqrt{t^{2} - z^{2}}=const$ in general does not correspond
to any isotherm, and a linear transverse rapidity profile is
questionable. However, based on estimates of corresponding
quantities in viable scenarios of hydrodynamic evolution and results
of correspondingly modified blast-wave model calculations (see,
e.g., Ref. \cite{Star-0} for blast-wave model calculations with
different velocity profiles), one can conclude that the spectra
discussed are little affected by the shape of the flow velocity
profile and by deviations of the chemical freeze-out isotherm from
the $\tau=const$ hypersurface. We  hence keep these assumptions
unchanged. Actually, the corresponding hypersurface and velocity
profile depend on the whole dynamical history of the
hydrodynamically expanding system - initial conditions, equation of
state, etc. - and can be calculated in a true dynamical model only.
It will be the subject of  follow-up work.

Another assumption, namely, the on-mass shell approximation of the
distribution function, may be questioned for finite-width particles
such as $J/\psi$, although the free width of the $J/\psi$ is less
than 100 keV. The main reason for the higher effective temperature
of $J/\psi$ spectra in the low-momentum region (see Fig. \ref{fig2})
 compared with the effective temperatures of $\phi$ and $\Omega$
spectra in the same region (see Fig. \ref{figg1}) is the larger mass
of the $J/\psi$ particle. An explanation of this fact can be based
on the simple approximation of the integrand in Eq. (\ref{1}). Then
one may surmise that  an appropriate off-mass-shell hydroinspired
parametrization of $J/\psi$ spectra that includes low invariant mass
contributions could improve the description of the low-momentum part
of the spectra. Most naturally, such a parametrization - namely, a
Breit-Wigner formula for the spectral function instead of the
$\delta$ function approximation - appears because of the finite
width of $J/\psi$ meson. However, the corresponding calculations of
the  $J/\psi$ spectra with the same hydrodynamical  parameters of
the blast-wave model as for $\Omega$ and $\phi$ particles  do not
result in a significant improvement in the description of $J/\psi$
spectra if experimental conditions are properly taken into account.
Namely, as  reported by the PHENIX Collaboration \cite{Phenix}, the
$J/\psi$ spectra at midrapidity are measured by counting lepton
pairs in the invariant mass range $2.9$ GeV $\leq M \leq $$ 3.3$
GeV, and such a narrow invariant mass window accompanied by the
rather low decay width, $\Gamma_{\text{tot}}=93.2 \times 10^{-6}$
GeV, does not result in noticeable deviation from on-mass shell
calculations.

The next assumption, namely, a neglect of flow fluctuations in the
blast-wave model, Eq. (\ref{01}), is also debatable. Owing  to the
finite size of the systems, large fluctuations are expected in the
initial stage of the nuclear collisions, even for a fixed impact
parameter, and various event generators do show such effects.  While
a very early thermalization in relativistic heavy-ion collisions is
doubtful now \cite{Sin} (for review see also Ref. \cite{therm} and
references therein), the initial inhomogeneities can result in
fluctuating initial conditions for the subsequent hydrodynamical
expansion. Because  the hydrodynamical equations are nonlinear, the
event average of any hydrodynamical  parameter (in other words,
average over solutions of hydrodynamical equations) is quite
different from that for a smooth initial configuration and results
in large differences in spectra from hydrodynamical calculations
with averaged initial conditions. This fact  was pointed out and
explicitly demonstrated in event-by-event hydrodynamic simulations
based on smoothed particle hydrodynamics code (see Ref. \cite{Hama1}
for review and Ref. \cite{Hama2} for recent results). Fluctuations
of transverse flows in noncental collisions also are considered  in
Ref. \cite{Ollit}.

\begin{figure}[h]
\includegraphics[scale=0.7]{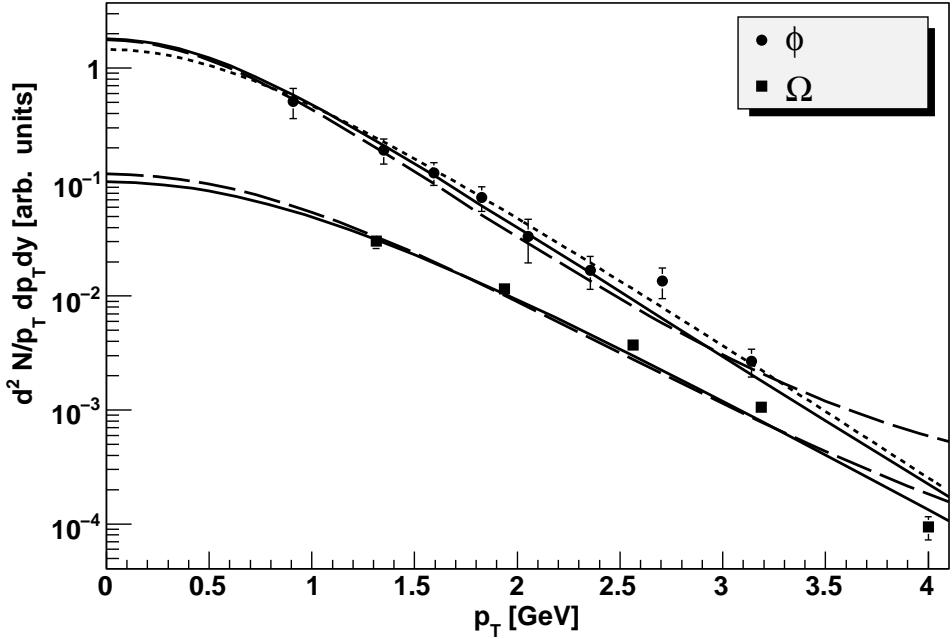}
\caption{\label{fig3} The same as  Fig.  \ref{figg1}, but with the
blast-wave model fit accounting for flow fluctuations, Eq.
(\ref{3}). The corresponding   results are indicated by  solid lines
(flat distribution in  hydrodynamical velocities) and dashed lines
(Gaussian distribution in  hydrodynamical velocities).}
\end{figure}

\begin{figure}[h]
\includegraphics[scale=0.7]{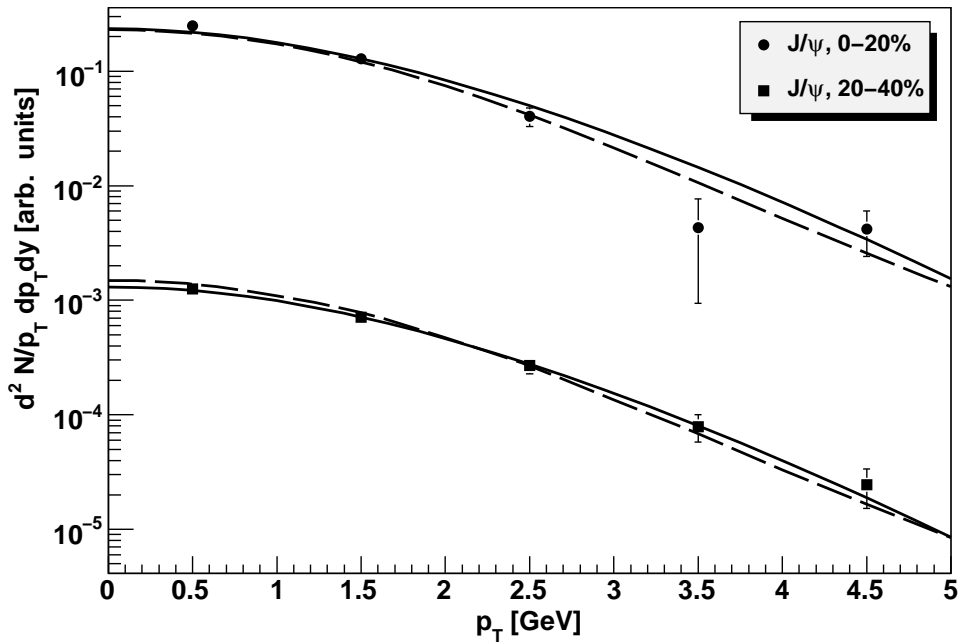}
\caption{\label{fig4} The same as  Fig.  \ref{fig2} but with the
blast-wave model fit accounting for flow fluctuations, Eq.
 (\ref{3}). The corresponding   results are indicated
by  solid lines (flat distribution in  hydrodynamical velocities)
and dashed lines (Gaussian distribution in hydrodynamical
velocities).}
\end{figure}

Therefore, the collective flow fluctuations are rather natural and
can have some effect on the particle momentum spectra. To study it,
let us consider a blast-wave model with bulk matter flow
fluctuations. The transverse spectra of the blast-wave model
averaged over ensemble of the fluctuations  are assumed to have the
form:
\begin{eqnarray}
 \frac{dN}{p_{T}dp_{T}dy}\propto m_{T}
\int_{\alpha_{\text{min}}}^{\alpha_{\text{max}}}d \alpha G(\alpha)
\int_{0}^{1} dxx I_{0}\left( \frac{p_{T}\sinh (\alpha x)}{T}\right)
K_{1}\left( \frac{m_{T}\cosh (\alpha x)}{T}\right). \label{3}
\end{eqnarray}
We consider here two simple and, in some sense, opposite cases of
the  distribution of $\alpha$. First, we assume a distribution that
is flat in $\alpha$, that is,  $G(\alpha)=1$, with the lower and
upper limits $\alpha_{\text{min}}$ and $\alpha_{\text{max}}$
considered to be  free fit parameters. Second, we assume a Gaussian
form for the distribution of hydrodynamical velocities. With
$v_{\text{max}}=\tanh \alpha$, we obtain: $G(\alpha)=\exp \left(-
\frac{(\tanh \alpha - \tanh \alpha^{0})^2}{\delta^{2}} \right)\equiv
\exp \left(- \frac{(v_{\text{max}} -
v^0_{\text{max}})^2}{\delta^{2}} \right)$, with
$\alpha_{\text{min}}=0$ ($v^{\text{min}}_{\text{max}}=0$) and
$\alpha_{\text{max}}=\infty$ ($v^{\text{max}}_{\text{max}}=1$). Note
that, for such a distribution, there is no explicit cutoff of high
$\alpha$ contributions; the convergence of the integral takes place
because the main contribution at high $\alpha$ happens at small
$x\sim 1/\alpha$, leading to a cutoff factor $1/\alpha^2$. The
results for $G(\alpha)=1$ with $\alpha_{\text{min}}=0.23$
($v^{\text{min}}_{\text{max}}\approx 0.23$) and
$\alpha_{\text{max}}=1.0$ ($v^{\text{max}}_{\text{max}} \approx
0.76$), and for $G(\alpha)=\exp \left(- \frac{(\tanh \alpha - \tanh
\alpha^{0})^2}{\delta^{2}} \right)$ with $\alpha^{0}=0.55$
($v_{\text{max}}^0\approx 0.5$) and $\delta=0.18$ are displayed in
Figs. \ref{fig3} and \ref{fig4} for $\Omega$, $\phi$, and $J/\psi$
transverse momentum spectra, respectively. One can see from these
figures that the blast-wave model supplemented with appropriate
fluctuations of the hydrodynamical flow yields a significantly
improved description of the shapes of transverse momentum
distributions for $\phi$, $\Omega$, and $J/\psi$ particles.

 Another assumption used in Eq. (\ref{01}), namely, that of
 homogeneous particle number density across the system at the isotherm
 hypersurface, is rather natural for bulk-matter particles consisting of
 quarks that are produced in QGP, as well as in the initial
 collisions, but can be questioned for heavy rare particles like $J/\psi$. The
 reasons are the following. Because the charm quarks are produced in primary
 binary nucleon-nucleon collisions only and there is no charm quark production
 during the evolution of the system, the initial distribution of charm quarks
 follows the distribution of initial binary collisions. Then the charm quark
 density can be more strongly peaked in the center of the system than the
 density of light-flavor and strange quarks that initially follow the
 participant density distribution.\footnote{It was recently demonstrated
   \cite{Girigori} that a difference in the initial distribution of
   charm quarks
   compared to the light flavors survives the hydrodynamical evolution and is
   still significant at decoupling, thereby resulting  in the reduction of the mean
   transverse momentum of the $D$-meson spectra.} Hence the effective
 transverse size of a volume occupied by $J/\psi$ mesons at the hadronization
 hypersurface can be less than that for bulk-matter particles.

For simplicity, we consider  the particle number density of $J/\psi$
in two extreme scenarios: first, the $J/\psi$ mesons are assumed to
follow the density distribution of bulk-matter particles, as  in
fact considered in the previous section, and second, we assume that
the spatial extension  of $J/\psi$ in the transverse direction
coincides with the initial spatial distribution of charm quarks
determined by the geometry of binary nucleon-nucleon collisions. The
latter is examined here.

The initial spatial distribution of charm quarks in the transverse
direction as determined by the transverse distribution of binary
nucleon-nucleon collisions in \textit{Au}+\textit{Au} collisions,
and at impact parameter $b=0$, which we assume  here for simplicity,
is
\begin{eqnarray}
n (r) \propto T_{A}^{2}(r), \label{d1}
\end{eqnarray}
where the nuclear thickness distribution $T_{A}(r)$ is obtained from
a Woods-Saxon distribution,
\begin{eqnarray}
T_{A}(r)=\int_{-\infty}^{+\infty}dz \rho \left ( \sqrt{r^{2}
+z^{2}} \right). \label{d2}
\end{eqnarray}
The Woods-Saxon distribution of  nucleons in the \textit{Au} nucleus
is defined by the formula
\begin{eqnarray}
\rho (\sqrt{r^{2} +z^{2}})= \rho_{0} \left [ 1+  \exp
\left(\frac{\sqrt{r^{2} +z^{2}}-c}{a} \right )
\right]^{-1}\label{d3}
\end{eqnarray}
with $c=6.38$ fm, $a=0.535$ fm \cite{table}, and $\rho_{0}$ given by
the normalization condition.

\begin{figure}[h]
\includegraphics[scale=0.7]{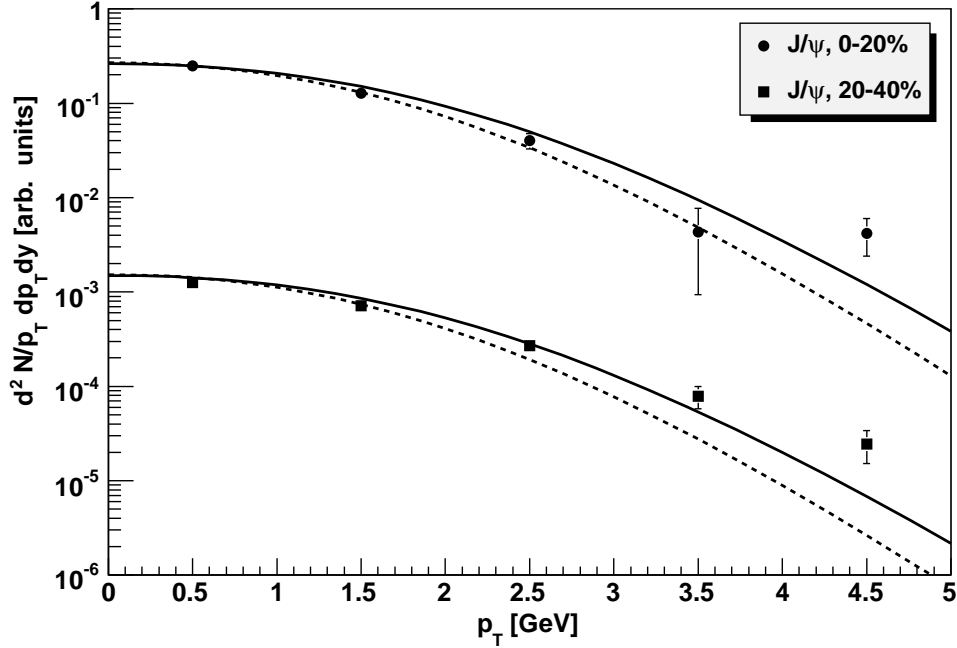}
\caption{\label{fig5} The same as  Fig.  \ref{fig2}, but with the
blast-wave model fit accounting for inhomogeneity of the $J/\psi$
distribution, Eq. (\ref{d4}), with $R=7$ fm (solid lines) and $R=8$
fm (dashed lines). }
\end{figure}

The transverse momentum spectrum of the blast-wave model of
$J/\psi$ distributed according to this initial spatial distribution
of charm quarks  is
\begin{eqnarray}
 \frac{dN}{p_{T}dp_{T}dy}\propto m_{T}
\int_{0}^{1} xdx  T_{A}^{2}(xR)I_{0}\left( \frac{p_{T}\sinh (\alpha
x)}{T}\right) K_{1}\left( \frac{m_{T}\cosh (\alpha x)}{T}\right).
\label{d4}
\end{eqnarray}
Then,  before calculating the  transverse momentum spectra of
$J/\psi$ by means of Eq. (\ref{d4}), one needs to define the
transverse size of the bulk-matter distribution $R$. We estimate
this value to be between $7$ and $8$ fm,  and  Fig. \ref{fig5}
presents the $J/\psi$ spectrum calculated with the hydroinspired
parametrization, Eq.  (\ref{d4}), for both $R$ values. A good
description  of the spectrum at low $p_T$ spectra is obtained, this
is not  surprising because, as  noted at the end of the previous
section, to fit the low-$p_T$ part of the $J/\psi$ spectrum one
needs to utilize, for $J/\psi$ mesons, a lower value of the maximum
transverse velocity than for bulk-matter particles at the same
isotherm.

The last assumption of the blast-wave model, Eq. (\ref{01}),
discussed here is the neglect of the resonance feed-down. While the
measured inclusive particle spectra in general contain contributions
from resonance decays, for the spectra discussed in this article the
resonance feed-down is rather small and one can expect that
generalization of the blast-wave model to include the resonance
feed-down does not seriously influence  the fitted
parameters.\footnote{See also Ref. \cite{Star-0} where it was
demonstrated
  that resonance decays have no significant effect on the kinetic freeze-out
  parameters extracted by means of the blast-wave model.} In any case,
feed-down contributions to $J/\psi$ spectra will, except for LHC
energies where feed-down from \textit{B}-meson decay is likely
significant, coming from excited charmonia. Although this
contribution can be up to 40 \% \cite{Satz2}, we expect little
modification of the spectral shapes,  as the charmonia masses are
all very large compared to masses of hadrons made of light and
strange quarks.  A comprehensive treatment of resonance feed-down
within a dynamical hydrokinetic model of the evolution of the
fireball and of particle production will be provided in a
forthcoming work.

\section{Results and discussion}

We have analyzed central  $J/\psi$, $\phi$, and $\Omega$ transverse
momentum spectra at midrapidity within a blast-wave model and
demonstrated that, while the high-$p_T$ part of the spectrum of
$J/\psi$ is well described by fit parameters obtained for a
description of $\phi$ and $\Omega$  spectra, the low-$p_{T}$
experimental transverse momentum spectrum of $J/\psi$ exhibits an
effective temperature lower  than that calculated with the fit
parameters for $\phi$ and $\Omega$ spectra. In our opinion, the
reason for this  is grounded in some specific assumptions of the
blast-wave model rather than in significant nonthermal contributions
to $J/\psi$ spectra. To demonstrate this   we have developed and
presented  a generalized blast-wave model.

First,  we have demonstrated that a blast-wave model modified to
account for collective flow fluctuations results in good agreement
with data over a wide $p_T$ region for $J/\psi$, $\phi$ and $\Omega$
transverse momentum spectra. However, good agreement with $J/\psi$
data  is reached  only if rather high flow fluctuations take place
near the chemical freeze-out isotherm. Note that in a very recent
paper \cite{Bron} it is estimated  that much lower flow
fluctuations, following from the fluctuations of the transverse size
of the initial source, are consistent  with recently observed
\cite{Star-fluct}  event-by-event fluctuations of the average
transverse momentum of bulk-matter particles in
\textit{Au}+\textit{Au} collisions at RHIC energies.\footnote{It is
worthy of note, also, that the magnitude   of event-by-event
elliptic flow fluctuations in  \textit{Au}+\textit{Au} collisions at
$\sqrt{s_{NN}}=200$ GeV \cite{Phobos} was found to be in agreement
with predictions based on spatial fluctuations of the participating
nucleons in the initial nuclear overlap region.} If this result is
confirmed, it rules out the interpretation that the low-$p_T$
peculiarities of $J/\psi$ transverse momentum spectra are a result
of  hydrodynamic flow fluctuations.

Another generalization of the blast-wave model considered in this
article  is based on the fact that, because charm quarks initially
follow the density distribution of primary binary nucleon-nucleon
collisions, the shape of $J/\psi$ density at freeze-out differs from
the particle density of uniform bulk matter because of a higher
concentration of charm quarks in the center of the system at the
hadronization hypersurface. The spectrum of $J/\psi$, calculated
with the density distribution of primary binary collisions and with
a conservative estimate of the transverse size of the volume
occupied by the bulk matter, demonstrates a fairly good agreement
with $0\%-20\%$ $J/\psi$ transverse momentum data until $p_{T}=3.5$
GeV, and the curve calculated is below the data points for higher
$p_{T}$. Note that high-$p_T$ $J/\psi$ can be emitted mostly by
corona and can be, in fact, the result of charmonium production in
nucleon-nucleon collisions (see Ref. \cite{PBM2}). Because the
fraction of participating nucleons contained in the corona region
increases when the centrality decreases, in noncentral collisions
one can expect a higher effective temperature of the $J/\psi$
spectra than in central collisions; indeed  Fig. \ref{fig5} shows
this effect.\footnote{The relatively high
  corona contribution is, perhaps, the reason why the $D^0$-meson spectra measured
  by the STAR Collaboration in the $0\%-80\%$ centrality bin \cite{Star-d} can
  be described by the relatively high radial flow velocity at the surface,
  $0.6$, and the freeze-out temperature, $170$ MeV  (see Ref.  \cite{Das}).}

 Finally, we conclude that the present data at the top RHIC
energy are compatible with the picture in which $J/\psi$, $\phi$,
and $\Omega$ momentum distributions are frozen simultaneously with
chemical composition at $T=160$ MeV. We  have demonstrated that a
consistent description of the $\phi$, $\Omega$, and $J/\psi$
transverse momentum spectra can be obtained within a
 hydroinspired parametrization of common
 freeze-out for these particles
 if nonhomogeneity of the charmonium
distribution and/or significant flow fluctuations are assumed.

\begin{acknowledgments}
 This work was supported  by  the Bilateral Grant  DLR (Germany)
- MESU (Ukraine) for the UKR 06/008 Project, Agreement  No.
M/26-2008,  and   by  a Ukrainian-French grant (DNIPRO Program),
Agreement with MESU No. M/4-2009.
\end{acknowledgments}

\end{document}